\documentclass[12pt]{iopart}
\usepackage{graphicx} 
\begin{document}

\title[Initial Conditions]
{Ekpyrosis and inflationary dynamics in heavy ion collisions: the role of quantum fluctuations}

\author{Kevin Dusling$^1$, Fran\c cois Gelis$^2$, and Raju Venugopalan$^{3}$}

\address{$^1$  Physics Department, North Carolina State University,
   Raleigh, NC 27695, USA}
\address{$^2$ Institut de Physique Th\'eorique (URA 2306 du CNRS),
  CEA/DSM/Saclay,  91191, Gif-sur-Yvette Cedex, France}
\address{$^3$ Physics Department, Bldg. 510A, Brookhaven National Laboratory,
   Upton, NY 11973, USA}
\ead{raju@bnl.gov}
\begin{abstract}

We summarize recent significant progress in the development of a first-principles formalism to 
describe the formation and evolution of matter in very high energy heavy ion collisions. The key role of quantum fluctuations both before and after a collision is emphasized. Systematic computations are now feasible to address early time dynamics essential for quantifying properties of strongly interacting quark-gluon matter. 
\end{abstract}


\section{Introduction}

The evolution of matter produced in the ``little bang" of heavy ion collisions, to borrow a 
term from a recent cosmological model~\cite{Khoury:2001wf}, is a process of Ekpyrosis\footnote{A term from Stoic cosmogony wherein the universe is destroyed in a fiery conflagration and subsequently reconstituted at regular intervals. Heavy ion collisions containing billions of events repeatedly undergo this process and are therefore closer in spirit to the Ekpyrotic scenario than the 
origin of the universe in an initial singularity.}, in which highly occupied coherent gluon states in the incoming nuclei are excited, producing very hot and dense matter shortly after the collision. The analog of the decaying inflaton field in this conflagration is a highly occupied Glasma~\cite{Kovner:1995ts,Krasnitz:1998ns,Lappi:2006fp} field. Quantum fluctuations play an essential role: a dominant class of these are responsible for the energy evolution of the Color Glass Condensate (CGC) nuclear states before the collision. The other dominant fluctuations grow exponentially with the square root of the proper time after the collision and likely isotropize matter in the Glasma. The mechanism of isotropization and possible thermalization bears close analogy to the inflationary dynamics of ``pre-heating" in the early universe~\cite{Polarski:1995jg,Kofman:1997yn,Micha:2004bv}. We will outline here recent work quantifying the role of quantum fluctuations in the early time dynamics in heavy ion collisions 

\section{Factorization and resummation of quantum fluctuations}

The treatment of high energy nuclear collisions in quantum field theory is motivated by the natural separation of scales between large $x$ static color sources $\rho_{1,2}^a$ in the nuclear wavefunctions and small $x$ dynamical gauge fields $A^{\mu,a}$. The evolution of this separation of scales leads to a Wilsonian renormalization group description formalized as the CGC  effective field theory~\cite{Gelis:2010nm}. At high energies, the color sources are strong ($\rho_{1,2}^a \sim 1/g$, where $g\ll 1$ is the QCD coupling) and color fields become time dependent immediately after the collision. Inclusive quantities, computed in the presence of these time dependent color fields\footnote{Examples of similar dynamics include Schwinger's mechanism for electron-positron pair production in strong QED fields and Hawking radiation in the vicinity of the event horizon of a black hole.}, are expressed in terms of retarded propagators, thereby allowing 
real time computations with initial data at negative infinity~\cite{Gelis:2006yv,Gelis:2006cr}. 

Before the collision, the effective ground states of the nuclei are coherent states described as classical fields with occupation numbers $O(1/g^2)$. In QCD, a classical description of Ekpyrosis is obtained by solving  Yang-Mills equations with static light front color  sources~\cite{Kovner:1995ts,Krasnitz:1999wc,Lappi:2003bi}. At this classical level,  energy dependence is introduced ``by hand" via the saturation scale $Q_S$, which is closely related to the variance of the  Gaussian random distribution of color charges in the MV model~\cite{McLerran:1993ni,McLerran:1993ka} of high nuclear wavefunctions. The energy dependence arises at next-to-leading (NLO) order from quantum fluctuations (of relative strength $g$) about the classical background of the nuclei. These naively sub-leading contributions are enhanced by logarithms, which give $\alpha_S\ln(x_{1,2}) \sim 1$ for small $x_{1,2}$ and have to be resummed to all orders in perturbation theory. In the strong field regime, there is an additional resummation $(g\rho_{1,2})^n$ at each order in the $\alpha_S$ expansion. These respectively radiative and multiple scattering contributions are generated by the JIMWLK renormalization group (RG) equation, ${\partial W[\rho_{1,2}]\over \partial \ln(x_{1,2})} = {\cal H}_{1,2}\; W[\rho_{1,2}]$, where $W[\rho_{1,2}]$ are gauge invariant weight functionals for the distributions of sources in the two nuclei, and ${\cal H}_{1,2}$ is the corresponding JIMWLK Hamiltonian~\cite{JalilianMarian:1997jx,Iancu:2000hn}.

An important consideration is to show that quantum fluctuations from the two nuclei don't talk to each other before the collision. Albeit required by causality, it is by no means assured in a weak coupling treatment; we are able to show formally that factorization of the contributions of the weight functionals $W[\rho_{1,2}]$ to inclusive quantities is obtained at leading logs in $x$ accuracy~\cite{Gelis:2008rw,Gelis:2008ad,Gelis:2008sz}. These quantum modes are boost-invariant $p^\eta=0$ modes, where $p^\eta$ is Fourier conjugate to the space-time rapidity. 

In the process of Ekpyrosis, nuclear coherence is lost and $p^\eta\neq 0$ modes are generated. These modes are generically unstable~\cite{Romatschke:2005pm,Romatschke:2006nk,Romatschke:2005ag,Fukushima:2011nq}, and grow in an expanding system as $(\alpha_S\exp(2\sqrt{\mu \tau}))^n$, with $\mu \sim Q_S$,  where $n$ denotes the order in perturbation theory beyond the classical 
leading order contribution.  These ``leading instabilities" are comparable to the background field at $\tau\sim 1/Q_S$ and have to be resummed to all orders, leading to qualitatively different behavior. 

This early time behavior has a robust analogy to ``pre-heating" dynamics in inflation~\cite{Polarski:1995jg,Kofman:1997yn,Micha:2004bv} where quantum fluctuations, enhanced by parametric resonance with the background Inflaton field, are conjectured to play an important role in ``turbulent thermalization" of the early universe. In particular,  quantum evolution of such a system can be represented by the stochastic average over repeated evolution of the inflaton field to which quantum seeds drawn from a Gaussian distributed ensemble of fluctuations are added~\cite{Son:1996zs,Khlebnikov:1996mc}. As we shall discuss, this formulations bears an exact analogy to the treatment of the problem in QCD~\cite{Dusling:2011rz}.

\section{Results and Outlook}

After factorization and resummation of leading logs in $x_{1,2}$ and leading instabilities $\alpha_S\exp(2\sqrt{\mu \tau})$, the energy-momentum tensor is expressed as 
\begin{equation}
\!\!\!\!\!\!\!\!\!\!\!\!\!\!\!\!\!\!\!\!\!\langle T^{\mu\nu}\rangle_{{\rm LLx + LInst.}} =\int [D\rho_1 D\rho_2]\;
W_{x_1}[\rho_1]\, W_{x_2}[\rho_2]  \int \!\! \big[{\cal D}\alpha\big]\,
F_0\big[\alpha\big]\; T_{_{\rm LO}}^{\mu\nu} [{\cal A}[\rho_1,\rho_2] + \alpha] (x)\; .
\label{eq:final-formula}
\end{equation}
The weight functionals $W_{x_{1,2}}[\rho_{1,2}]$ satisfy the
JIMWLK RG equation discussed previously. The
argument ${\cal A}\equiv ( A, E)$ denotes collectively
the components of the classical fields and their canonically
conjugate momenta on the initial proper time surface; analytical expressions for these are available at
$\tau=0^+$~\cite{Kovner:1995ts,Krasnitz:1998ns}.  The initial spectrum of fluctuations $F_0\big[\alpha\big]$, Gaussian in the quantum fluctuations $\alpha$,  has a variance given by the small fluctuation propagator in the Glasma background field as $\tau\rightarrow 0^+$. In practice, the path integral in $\alpha$ is determined by solving the classical Yang-Mills equations repeatedly with the initial conditions at $\tau=0^+$ given by
\begin{equation}
{\bf A}_{\rm init.} = {\cal A}_{\rm init.} + \int d\mu_{_K}\;\Big[c_{_K}\,a_{_K}^\mu(x)+c_{_K}^*\,a_{_K}^{\mu*}(x)\Big] \, .
\label{eq:quantum}
\end{equation}
Here ${\bf A}$ collectively denotes the quantum fields and their canonical conjugate momenta. The  coefficients $c_{_K}$, with $K$ collectively denoting the quantum numbers labeling the basis of solutions, are random Gaussian-distributed complex numbers given by
\begin{eqnarray}
\left<c_{_K}c_{_{K^\prime}}^*\right>=\frac{N_{_K}}{2} \delta_{_{KK^\prime}}
\;\; ,\;\;\left<c_{_K}c_{_{K^\prime}}\right>=\left<c_{_K}^*c_{_{K^\prime}}^*\right>=0 \; .
\label{eq:random2}
\end{eqnarray}
Explicit expressions for the small fluctuations and their conjugate momenta, denoted here by $a_K^\mu (x)$ were obtained in ref.~\cite{Dusling:2011rz}. The inner product of these solutions satisfies the orthogonality condition $(a_{_K}\big|a_{_{K^\prime}}\big)=N_{_K}\,\delta_{_{KK^\prime}}$ with the measure $d\mu_{_K}$ (a mix of integrals and discrete sums) that ensures $\int d\mu_{_K}\;N_{_K}\,\delta_{_{KK^\prime}}=1$. 

In ref.~\cite{Dusling:2011rz}, a numerical algorithm was outlined to compute eq.~(\ref{eq:final-formula}), thereby describing both Ekpyrosis and inflationary dynamics including essential leading quantum fluctuations. (We emphasize the formalism holds for any inclusive quantity in heavy-ion collisions including for instance parton energy loss and sphaleron transitions at early times.) We can now study how lumpy initial conditions for color charges $\rho_{1,2}^a$ in nuclear wavefunctions at the energy/rapidity of interest transform into the flow of matter.  The early universe analogy is becoming more robust experimentally with a ``WMAP-like" spectrum of spatial anisotropies now being resolved in heavy ion data~\cite{Sorensen:2011hm}.

We carried out an extensive study of the formalism outlined here for a scalar $\phi^4$ model which, among several  QCD-like features, has a spectrum of unstable quantum modes which are amplified by resonant interactions with the background field~\cite{Dusling:2010rm}. In $\phi^4$ (and other scale invariant theories), the amplitude of the field is inversely proportional to its period of oscillation. Slightly different amplitudes, corresponding to different quantum seeds, lead to differing oscillation periods; a stochastic average over these leads to decoherence in the evolution.  A striking consequence is hydrodynamic flow with an ideal equation of state. Eq.~(\ref{eq:quantum})  is a  realization of Berry's conjecture~\cite{Berry} which is believed to be necessary for thermalization~\cite{Srednicki} of a quantum system. Thermalization and onset of quasi-particle dynamics have been studied in the scalar theory; similar studies are feasible in the QCD framework of ref.~\cite{Dusling:2011rz}.

\section*{Acknowledgments}
R.V. and K.D are supported by the US Department of Energy under
DOE Contract DE-AC02-98CH10886 and  DE-FG02-03ER41260.

\end{document}